# Three-dimensional imaging of porcine joint tissues down to the subcellular level

Georg Schulz*[a,b], Andrea Barbero[c], Francine Wolf[c], Griffin Rodgers[a], Christine Tanner[a], Timm Weitkamp[d], Marcus Mumme[e], Marta Morawska[f], Daniel Beer[f], Bert Müller[a]

[a]Biomaterials Science Center, Department of Biomedical Engineering, University of Basel, Gewerbestrasse 14, 4123 Allschwil, Switzerland;
[b]Core Facility Micro- and Nanotomography, Department of Biomedical Engineering, University of Basel, Gewerbestrasse 14, 4123 Allschwil, Switzerland;
[c]Department of Biomedicine, Univeristy of Basel, Hebelstrasse 20, 4056 Basel, Switzerland,
[d]Synchrotron SOLEIL, L'Orme des Merisiers, 91190 Saint-Aubin, France;
[e]Regenerative Cartilage Surgery, University Hospital of Basel, Spitalstrasse 21, 4031 Basel,
[f]Advanced Osteotomy Tools, Wallstrasse 6, 4051 Basel, Switzerland

## ABSTRACT

Joint tissues consist of trabecular and cortical bone as well as calcified and hyaline cartilage, which presents a challenge for hard X-ray-based visualization on the sub-cellular level due to the wide range of local X-ray absorption values. The density of the calcified tissues requires rather high photon energy, which often leads to insufficient contrast within the cartilage and impedes the visualization of individual biological cells. Decalcification of the tissues reduces the total and local X-ray absorption values and allows for selecting a lower photon energy. Further contrast enhancement can be achieved by ethanol fixation and paraffin tissue embedding. In this study, we (i) searched for an appropriate visualization method to investigate lesions generated by a laser osteotome and (ii) visualized a decalcified porcine joint after ethanol fixation and subsequent paraffin embedding using laboratory- and synchrotron radiation-based microtomography. The experiments at the ANATOMIX beamline of Synchrotron SOLEIL were performed in off-axis scan mode with a pixel size of 1.3 µm. Individual cells in all layers of the joint could be made visible and the effect of ethanol fixation and paraffin embedding demonstrated.

## 1. INTRODUCTION

In the past decade, robot-guided lasers have become an established method to cut bones. Compared to conventional saws, they provide higher cutting precision and can lead to improved bone healing. It is reasonable to expect that in near future robot-guided lasers will be successfully employed to excavate damaged cartilage from joints. The present study aimed at the selection of an appropriate imaging method and the examination of a potential sample preparation protocol for investigating lesions generated by CARLO® (Cold Ablation Robot-assisted Laser Osteotome, Advanced Osteotomy Tools AG, Basel, Switzerland) [1], which is one of the world's first digital, orthopedic laser surgery platforms. The system is equipped with a 2940-nm Er:YAG laser targeting the water content in the tissue. The instantaneous evaporation is causing micro fractures and blasting out a certain volume depending on the energy setting. By this principle of focused laser energy deposition, the tissue is ablated without detectable subsequent thermal damage. Unlike the mechanical instruments currently used to perform osteotomy, laser-induced bone ablation does not create a smear layer on the osteotomy edges, resulting in a channeled scaffold that preserves the trabecular ridges, which allows the passage of cells to the site of injury, potentially benefiting bone healing [2]. Laser ablation has been also experimentally used on knee cartilage [3], for example as an alternative to microfracturing. However, the number of studies is limited and meaningful comparisons with current medical standards are largely missing [4]. In the current pilot investigation, a fresh domestic pig knee was selected.

---

*Send correspondence to G.S., e-mail: georg.schulz@unibas.ch; telephone: +41 61 207 54 37, https://www.bmc.unibas.ch

Microtomography was chosen as a method to help quantitatively compare the performance, accuracy, and tissue preservation of laser ablation versus manual professional curettage. In hospitals, the investigation of joints is daily routine. Bony tissues can be visualized with a spatial resolution down to a fraction of a millimeter by clinical computed tomography, whereas the soft tissues can be made visible by magnetic resonance imaging with a slightly inferior spatial resolution. These ranges of spatial resolution rule out investigations on the cellular level. Several microscopic studies of bone biopsies from humans and animals with resolution from the micron range [5-8] to the deep submicron domain [9] are available, but they mainly concentrate on the hard tissues. The visualization of knee joint specimens using hard X-ray microtomography is challenging, as it contains the rather soft hyaline cartilage and the much harder calcified cartilage and bony tissues. As the density of these materials differs by about a factor of two, it is difficult to find the proper X-ray photon energy to simultaneously visualize both tissue types. A photon energy suitable for bone will be too high for soft tissue components and thus will provide insufficient contrast within the cartilage. Conversely, a photon energy well suited to visualizing soft tissues will generate streaks and other artifacts owing to the absorbing bone [10]. One possible workaround is the use of contrast agents such as osmium tetroxide [11]. An alternative would be phase-contrast-based microtomography, because the signal is less dependent on the atomic number of the components than in conventional absorption contrast [12, 13].

The present study concentrates first on the macroscopic three-dimensional visualization of the entire porcine condyle including bony tissues and hyaline cartilage, to observe the performance of laser and manual ablation. The second step is the visualization of cylindrical biopsies on the cellular level without the use of any contrast agent. For theses tasks an advanced conventional microtomography system and a synchrotron-radiation microtomography setup were used.

## 2. MATERIALS AND METHODS

### 2.1 Tissue preparation

The cartilage ablation study was carried out on a domestic pig condyle specimen. Two ablations were performed with CARLO®: the shallow ablation (L-s) that should remove the hyaline cartilage and stop at the calcified cartilage and the deep ablation (L-d) that should also remove the calcified cartilage and stop at the cortical bone, see Fig. 1a. In both cases, the focal point of the laser was set above bone surface, to increase the width of the cut segment, resulting in smooth ablation surface. For comparison, an experienced orthopedic surgeon with specialization in cartilage repair performed a manual ablation of the cartilage (M) using a surgical curette. After the first laboratory microtomography experiments the joint was formalin fixated. After the fixation four osteochondral blocks (L-s, L-d, M and Ctrl) were extracted and decalcified for a period of 60 days using a 7% EDTA 30% sucrose solution (Sigma-Aldrich). Then two cylinders 6 mm in diameter were punched out of each of the four regions, cf. Fig. 1c. These osteochondral cylinders were then processed histologically, i.e. ethanol series (50% to 100%), applied to xylene and embedded in paraffin. Tissues were then sectioned (thickness 5 to 7 µm) and stained with Safranin-O / Fast Green to visualize the sulphated glycosaminoglycans (GAG) positive matrix and the collagenous matrix.

### 2.2 Laboratory-based microtomography

The laboratory-based tomography experiments were performed with a nanotom® m (Waygate Technologies, Wunstorf, Germany). With a distance of 60 cm between source and detector and a flat-panel detector with $3,072 \times 2,400$ pixels, the system allows visualizing specimens with diameters of up to 35 cm [13]. For substantially smaller objects, the nanofocus X-ray tube also provides a cellular resolution [14]. The scan was acquired with an acceleration voltage of 120 kV and a beam current of 100 µA. A 0.25-mm-thick copper filter was used to increase the mean photon energy. The effective pixel size was set to 25 µm. A set of 1,500 projections was acquired over 360° with an exposure time of 4 s at each rotation angle, resulting in a total scan time of 2 hours and 15 minutes. For the visualization of the decalcified cylinders after ethanol fixation and paraffin embedding a lower mean photon energy was used, i.e. an acceleration voltage of 90 kV and a beam current of 200 µA. With an effective pixel size of 5 µm, 1,440 projections were acquired along 360° with an exposure time of 6 s per projection.

### 2.3 Synchrotron radiation-based microtomography

The synchrotron radiation-based tomography experiments were performed at the ANATOMIX beamline of Synchrotron SOLEIL (Saint-Aubin, France) [15] using a filtered white beam with a mean photon energy of around 17 keV (undulator gap of 8.3 mm and a gold filter of 10 µm thickness). The data was detected by a scientific CMOS camera (Hamamatsu Orca Flash 4.0 V2, $2048 \times 2048$ pixels, 6.5 µm physical pixel size) equipped with a 20 µm-thick LuAG scintillator. Measurements were made at effective pixel sizes of 0.65 µm and 1.30 µm, obtained using microscope objectives (Mitutoyo, Kawasaki, Japan) with respective magnifications of 10× and 5×. A set of 5,900 projections was acquired over an angular range of 360° with off-axis rotation, which almost doubled the field of view in the horizontal dimensions. The exposure time was set to 50 ms, which filled half of the detector's dynamic range. The propagation distance was set to 50 mm for single-distance phase retrieval [16].

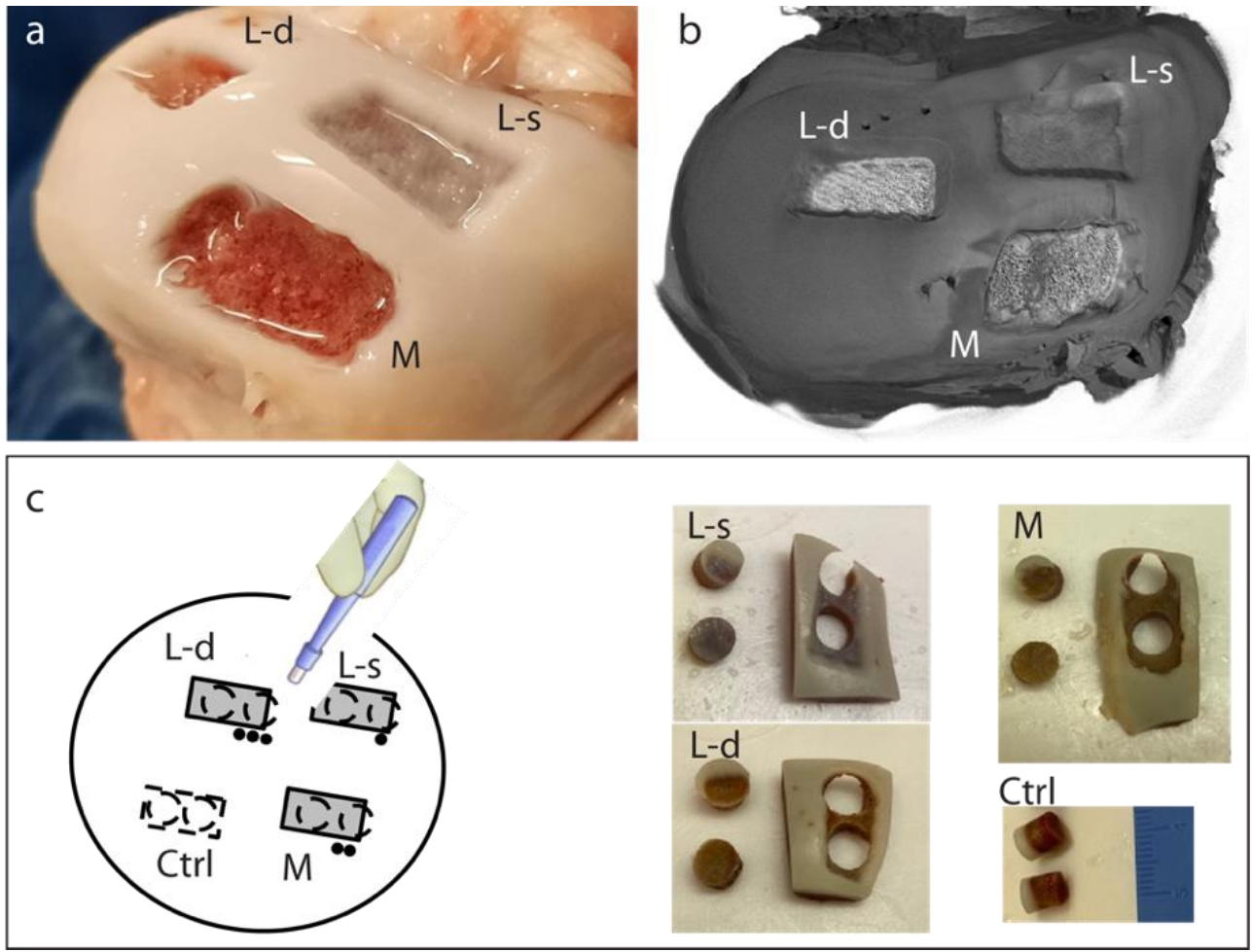

Figure 1. Preparation steps of the pig condyle. a: Fresh condyle after manual (M), deep laser (L-d) and shallow laser (L-s) ablations. b: Three-dimensional rendering of the porcine condyle on the basis of tomography data. c: Cylindrical osteochondral specimens after formalin fixation and decalcification of the three ablated regions plus a control, not ablated region (Ctrl).

## 3.  RESULTS AND DISCUSSION

Figure 2 shows renderings and orthogonal virtual cuts through the lesions of the joint before formalin fixation. As intended, the shallow laser cut only removed the hyaline cartilage and did not reach the calcified cartilage. The deep laser and the manual lesion treatment also removed some of the calcified cartilage and was even partially extended to the subchondral bone. The laser cut did not have a uniform depth over the lesion, cf. virtual cut in Fig. 2 bottom left, as the planning did not include information of the precise location of the cartilage. This variation can be corrected by incorporating optical coherence tomography measurements during the cutting procedure. Apart from this detail, the laser-ablated lesion was more precise compared to the manual one, in the sense that all surfaces, i.e. cartilage and bone, were flat and homogeneous.

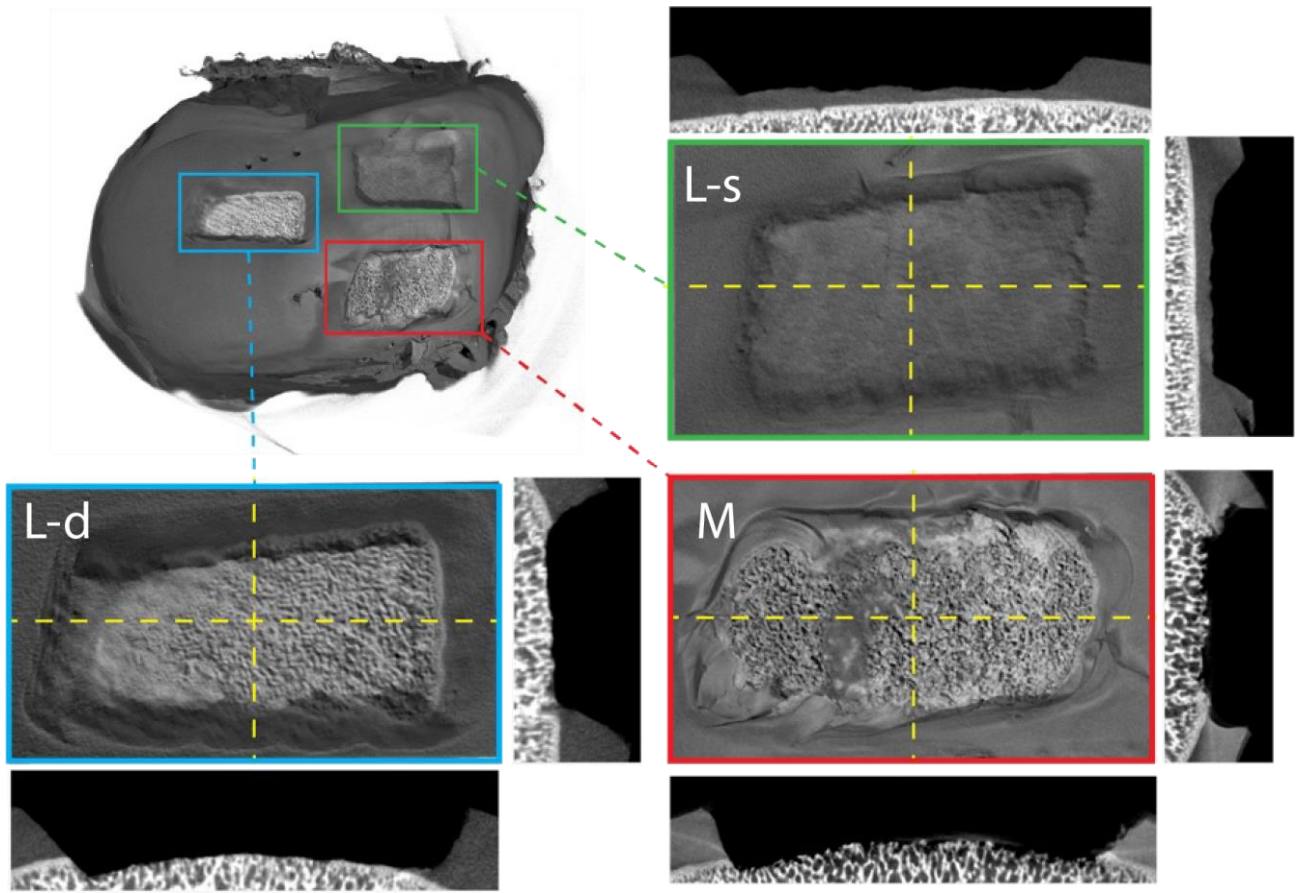

Figure 2. Three-dimensional renderings from the conventional microtomography data set of the three ablated regions and virtual cuts indicate the depth and the quality of the ablations. The three frames "L-s", "L-d", "M" are enlarged details of the top-view 3D rendering shown on the upper left. The other images are vertical sections (slices) through the tomographic volume taken along the dashed yellow lines.

Figure 3 shows histological slices of the control osteochondral cylinder after Safranin-O / Fast Green staining. This staining colors the GAG positive matrix in red and the collagenous matrix in bluish green. The histological slice reveals the microanatomy of bone and cartilage and illustrates the morphology of the superficial layer, the deep zone, the calcified cartilage and the bone. The images in Fig. 3 also show the challenges of histological sectioning and staining, i.e. partial tissue folding and formation of gaps.

With hard X-ray microtomography it is possible to correct these deformations and obtain additional information on the cellular level. It can be performed at one of the individual preparation steps (formalin fixation, decalcification, ethanol fixation, paraffin embedding) [17, 18]. Because of the challenges in simultaneous visualization of cartilage and bone [10], the imaging was performed after formalin fixation and decalcification. In order to investigate the differences between ethanol fixation and paraffin embedding on the macroscopic level, laboratory-based experiments were performed. Figure 4 shows virtual cuts through the tissues after ethanol fixation (left) and paraffin embedding (right). While the contrast in the tomography data after paraffin embedding is increased, the morphology of the specimen has drastically changed between the two preparation steps. The cartilage appears inhomogeneously deformed. Moreover, the tissue after ethanol fixation shows sufficient contrast (albeit inferior to imaging after paraffin embedding) to identify bone, cartilage, and even calcified cartilage, i.e. the dark layer between bone and cartilage.

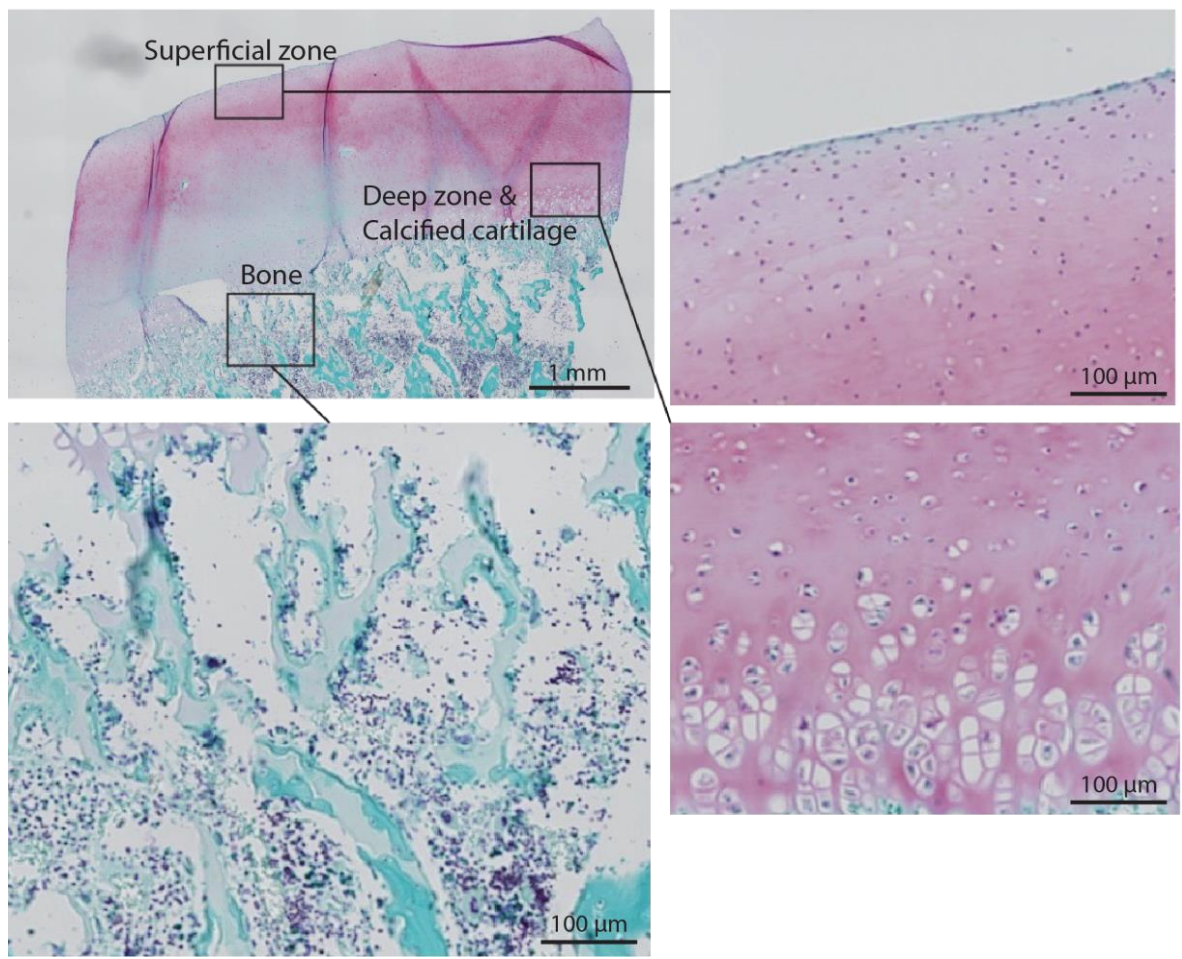

Figure 3. Optical micrographs of a histological slice of the decalcified, paraffin-embedded porcine osteochondral cylinders after Safranin-O / Fast Green staining. The magnified views show three selected regions of the tissues and illustrate their appearance on the microscopic level.

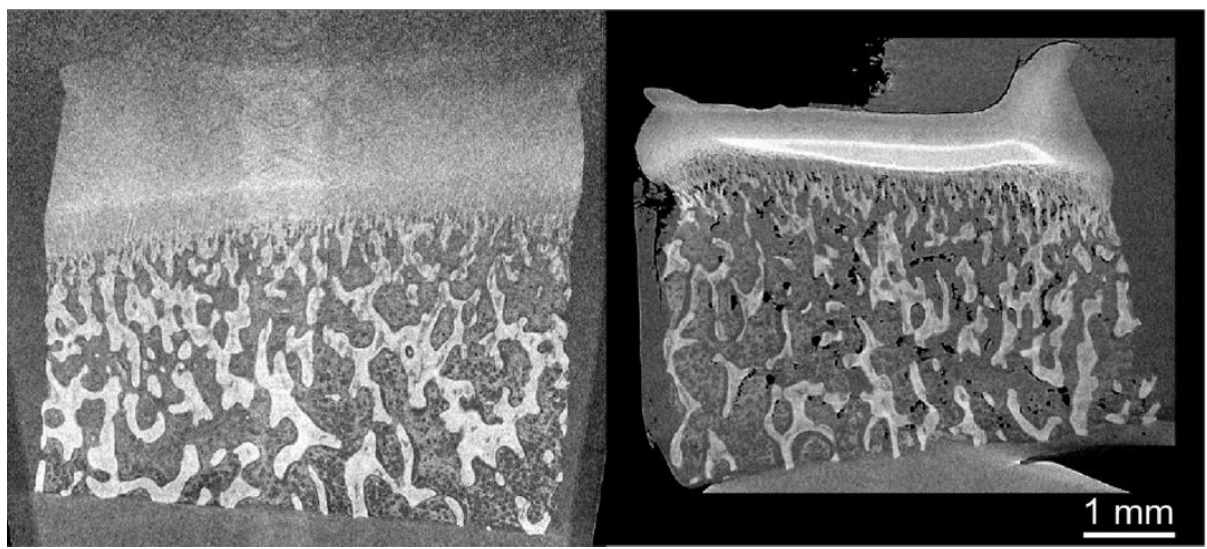

Figure 4. Laboratory-based microtomography slices through the ethanol-fixated (left) and paraffin-embedded (right) porcine osteochondral tissues indicating the macroscopic deformations induced by paraffin embedding.

For the comparison on the cellular level, synchrotron radiation-based microtomography was used. Figure 5 shows virtual cuts through the decalcified tissue after ethanol fixation (left) and paraffin embedding (right). The quality (contrast-to-noise, no deformation of the specimen) of the tomography data after ethanol fixation is obviously better. Paraffin embedding also induces artifacts in the morphology that are apparent in the images. Within the bony tissue, air inclusions are present. These inclusions are especially obstructive in the phase-contrast microtomography data sets, as they produce artifacts due to the phase shift difference between paraffin and air. Both data sets illustrate the morphology of the joint on the cellular level, including individual, specifically shaped chondrocytes.

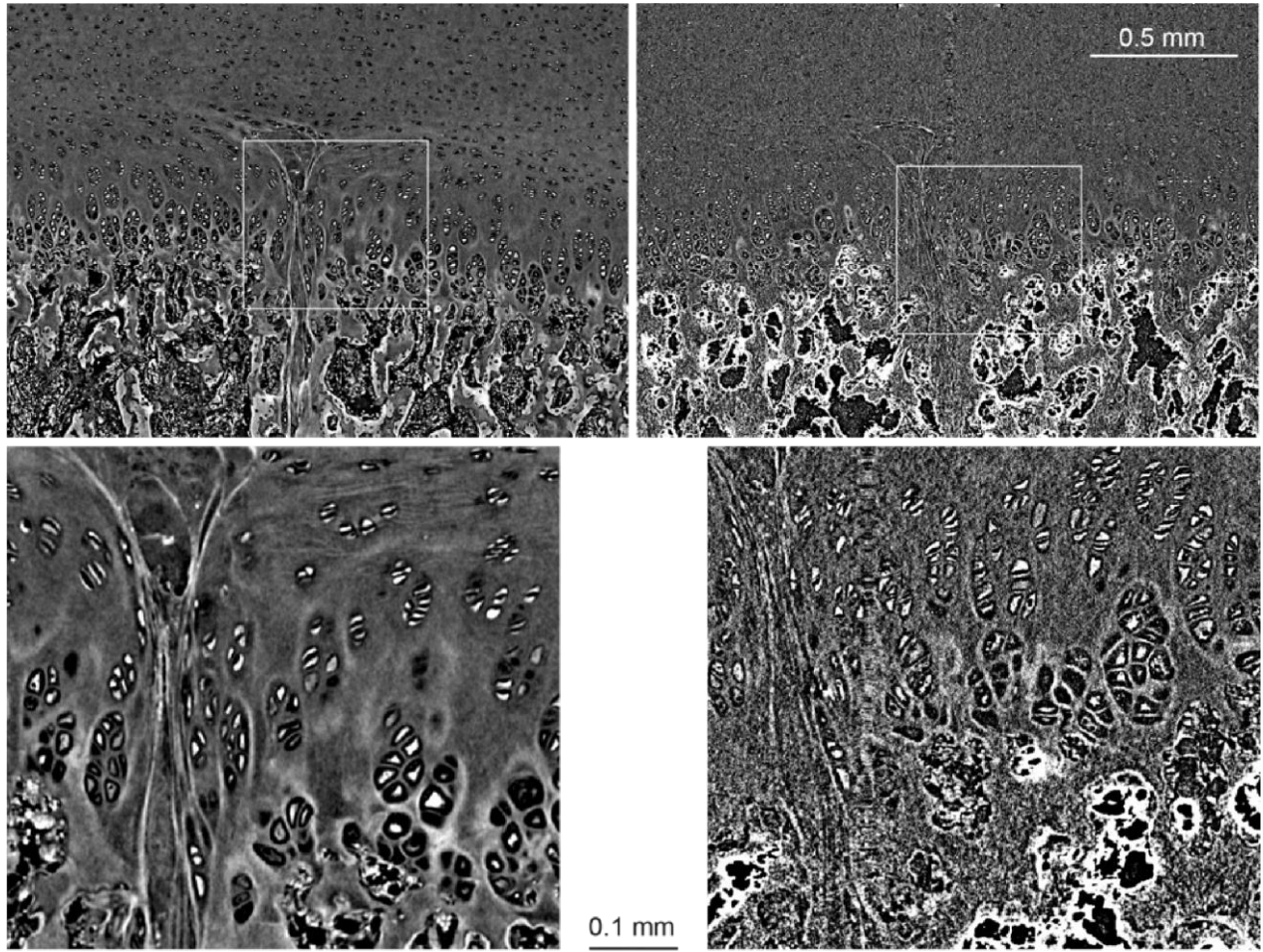

Figure 5. Synchrotron-radiation microtomography slices through the ethanol-fixated (left) and paraffin-embedded (right) porcine osteochondral tissues show the difference in contrast for the two preparation routes and artifacts in the paraffin data sets induced by air inclusions. The bright features are chondrocytes in lacunae.

The comparison between a data set acquired with an effective pixel size of 1.30 µm (Fig. 6 left) and one of 0.65 µm (Fig. 6 right) showed that the former (i.e., lower spatial resolution) is preferable for the investigated tissues. Similar anatomical features could be observed in both data. The advantage of using the larger pixels rather than the submicron pixels is that the volume imaged in a single scan is eight times larger. Figure 7 shows a three-dimensional rendering and three orthogonal cuts through the specimen, where the horizontal cut (red) was chosen through the cartilage.

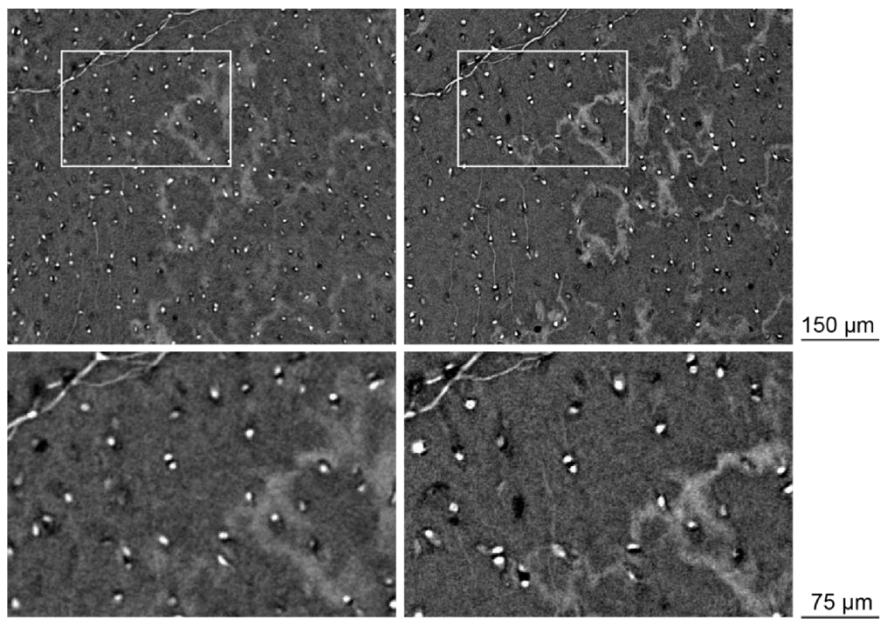

Figure 6. Synchrotron radiation-based microtomography slices through the specimen acquired with effective pixel sizes of 1.30 µm (left) and 0.65 µm (right) indicate minor differences in contrast and spatial resolution.

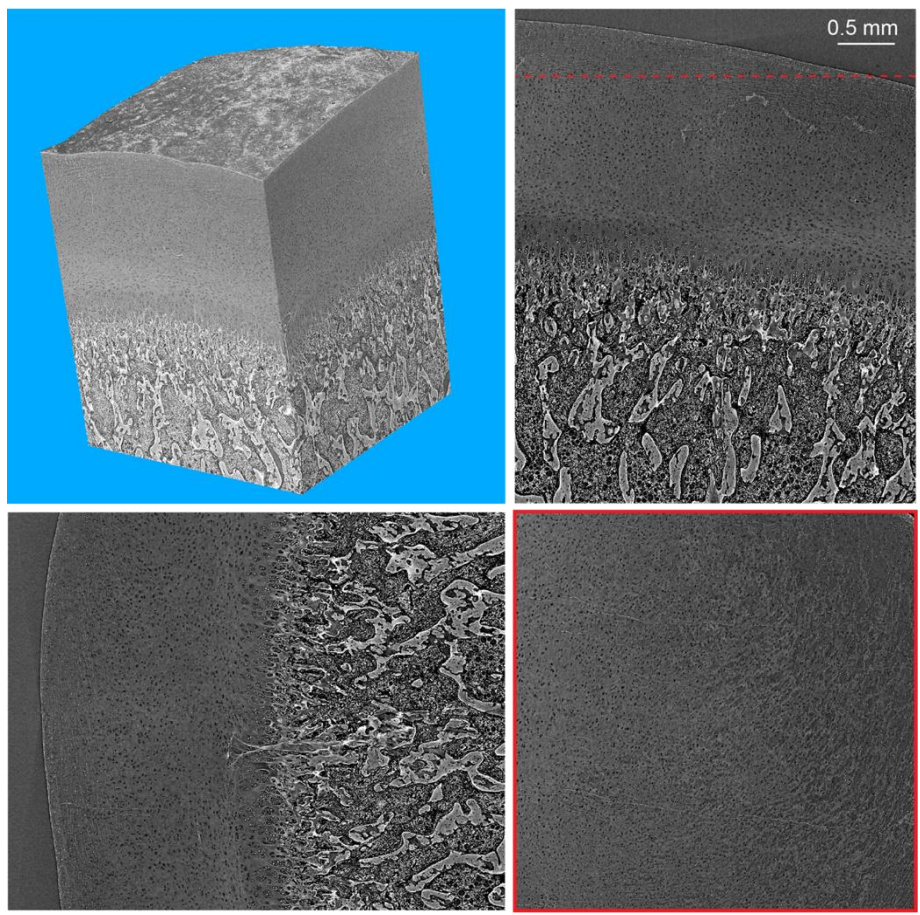

Figure 7. Three-dimensional rendering and orthogonal virtual cuts through the synchrotron radiation-based microtomography data (1.3 µm-wide voxels) of the ethanol-fixed porcine joint specimen showing the morphology with sub-cellular features.

A more detailed view of the virtual slice in Fig. 7 top right is shown in Fig. 8. All the regions could be identified: the *superficial zone* (blue) with the flat chondrocytes; the *intermediate zone* (green) with spheroid-shaped chondrocytes; the *deep zone* (yellow) with the chondrocytes aligned in columns; the *calcified cartilage* (magenta) with the chondrocyte clusters surrounded by calcified extracellular matrix; and the decalcified bone (white).

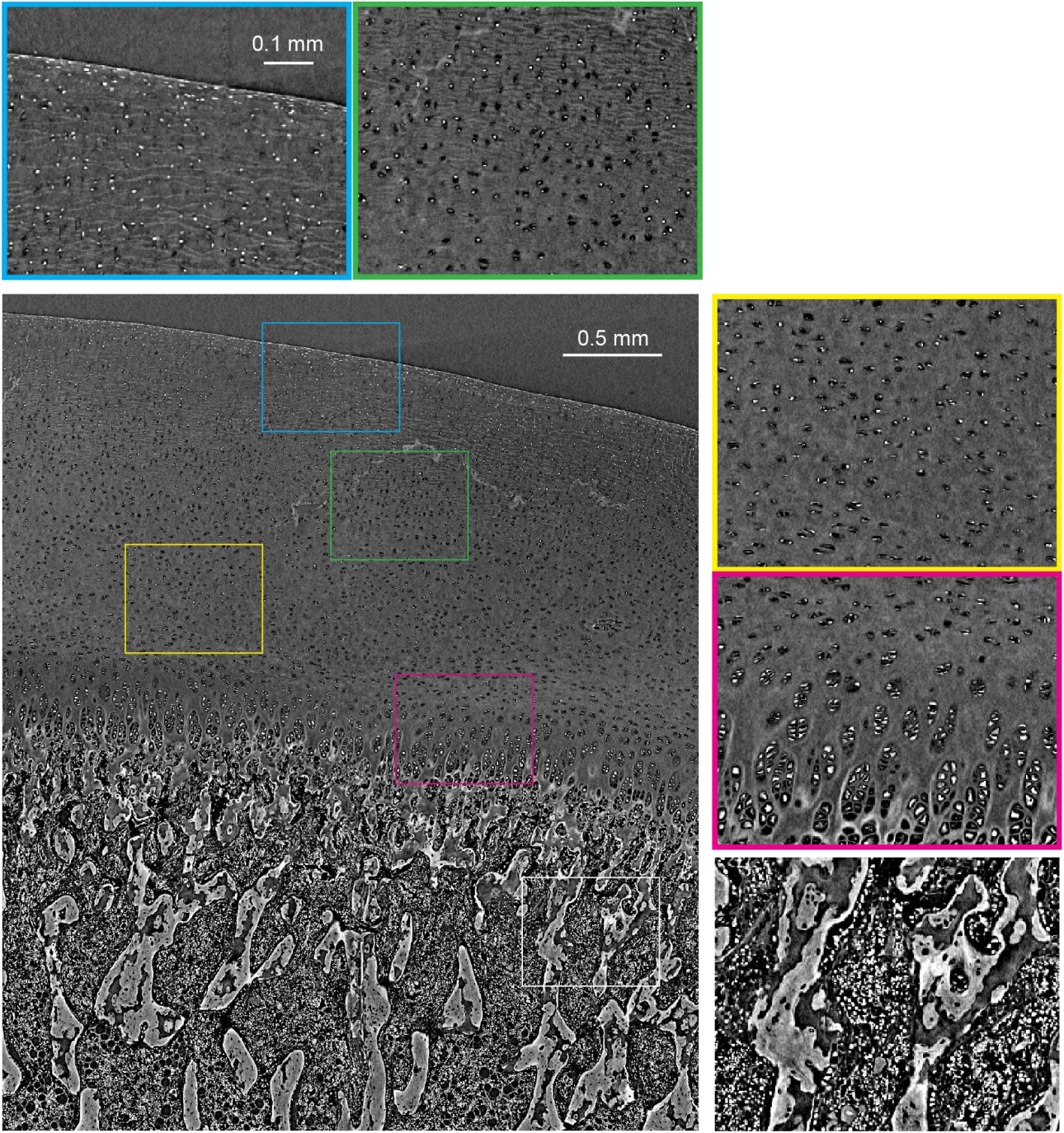

Figure 8. Synchrotron radiation-based microtomography slice of the ethanol-fixed porcine joint tissue including magnified views into the superficial zone (blue framed), the intermediate zone (green framed), the deep zone (yellow framed), the calcified cartilage (magenta framed) and the decalcified bone (white framed) showing sub-cellular anatomical features.

## 4. CONCLUSIONS

For the simultaneous visualization of cartilage, calcified cartilage and bony tissues of a knee using X-ray microtomography, the decalcification of the specimen was required. The contrast could be increased by ethanol fixation. Paraffin embedding is hardly beneficial, as it produces strong deformations of the cartilage and because it is very challenging to remove air inclusions within the trabeculae of the bone. Macroscopic investigations on the entire joints and parts of it can be performed with a laboratory-based tomography system. For cellular resolution single-distance phase-contrast microtomography at synchrotron radiation facility can be applied. Looking on the feature sizes within the porcine joint specimen, an effective pixel size of 1.30 µm is preferable, as it provides an eightfold larger volume compared to 0.65 µm-wide pixels with similar visual details. Other advantages of the lower resolved data set are lower dose and shorter scan duration to achieve equivalent contrast-to-noise ratio. A mapped ROI with the same volume content, includes an eight times smaller size of the data set. The lower resolved data set allowed for the detection of chondrocytes and consequently for the differentiation between the cartilage types. The described radiological examinations will be used in future studies to investigate the performance of emerging technologies in which the laser is combined with a real-time feedback loop to control the ablating process, i.e. using optical coherence tomography. Moreover, such radiological assessments could be performed to obtain a more accurate description — as compared to that provided by histological analyses — of the quality of biopsy samples collected from operated joints of patients or of implantable cartilage tissues manufactured for the treatment of patients.

## ACKNOWLEDGEMENTS

The authors acknowledge financial support of the Swiss National Science Foundation (SNSF) in the frame of the R'equip initiative (133802). ANATOMIX is an Equipment of Excellence (EQUIPEX) funded by the *Investments for the Future* program of the French National Research Agency (ANR), project *NanoimagesX*, grant no. ANR-11-EQPX-0031.